\documentclass[12pt,a4paper]{article}
\usepackage{amsmath,caption,bbm,longtable,paralist}
\usepackage[top=1.3in, bottom=1.3in, left=1.1in, right=1.1in]{geometry}

\DeclareCaptionStyle{italic}[justification=centering]{labelfont={bf},textfont={it},labelsep=colon}
\usepackage{graphicx,psfrag,epsf}
\usepackage{enumerate}
\usepackage{natbib}
\usepackage{url}
\usepackage{bbm, dsfont}
\usepackage{comment}
\usepackage{geometry}
\usepackage{float}
\usepackage[dvipsnames]{xcolor}
\usepackage{booktabs, subfig, bm, paralist,mathpazo,tikz,todonotes,longtable,microtype,dsfont,rotating} 
\usepackage[pdftex,colorlinks=true]{hyperref}
\definecolor{darkblue}{rgb}{0,0,.6}
\hypersetup{citecolor=darkblue,linkcolor=darkblue,urlcolor=darkblue}

\newcommand{\blind}{0}

\newcommand{\X}{\mathcal{X}}
\newcommand{\Y}{\mathcal{Y}}

\graphicspath{{plots/}}
\DeclareMathOperator*{\argmin}{\arg\!\min}
\newsavebox\CBox

\definecolor{a0}{rgb}{0.0, 0.5, 0.0}
\definecolor{bistre}{rgb}{0.24, 0.17, 0.12}
\definecolor{amethyst}{rgb}{0.6, 0.4, 0.8}
\definecolor{blue-violet}{rgb}{0.54, 0.17, 0.89}
\definecolor{Rcolor}{RGB}{150,160,190}
\definecolor{blush}{rgb}{0.87, 0.36, 0.51}
\definecolor{brightturquoise}{rgb}{0.03, 0.91, 0.87}
\definecolor{burntorange}{rgb}{0.8, 0.33, 0.0}

\usepackage{orcidlink}
\date{}
\AtBeginDocument{}
\newcommand{\Rlogo}{\protect\includegraphics[height=1.8ex,keepaspectratio]{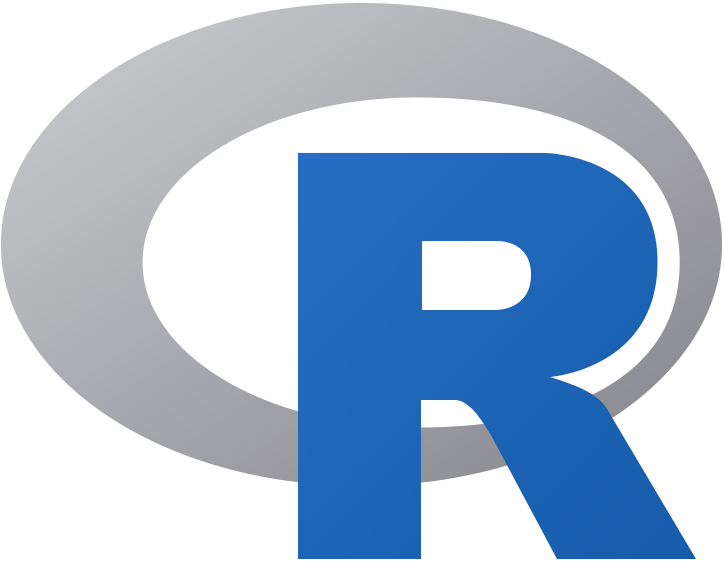}}

\begin{document}

\def\spacingset#1{\renewcommand{\baselinestretch}{#1}\small\normalsize} \spacingset{1}

\if0\blind
{
  \title{\bf Forecasting density-valued functional panel data}
  \author{\normalsize Cristian F. Jim\'enez-Var\'on \orcidlink{0000-0001-7471-3845}\thanks{Postal address:  Department of Mathematics. University of York, Heslington, York YO10 5DD, United Kingdom. E-mail: cristian.jimenezvaron@york.ac.uk} \\
   \normalsize Department of Mathematics\\
   \normalsize University of York\\
   \normalsize Department of Physics and Mathematics\\
  \normalsize Universidad Autónoma de Manizales \\
  \\
  \normalsize Ying Sun \orcidlink{0000-0001-6703-4270}   \hspace{.2cm}\\  
  \normalsize CEMSE Division, Statistics Program\\ 
  \normalsize King Abdullah University of Science and Technology \\ 
  \\
  \normalsize Han Lin Shang \orcidlink{0000-0003-1769-6430} \\
    \normalsize Department of Actuarial Studies and Business Analytics \\
    \normalsize Macquarie University
}
  \maketitle
} \fi

\if1\blind
{
\title{\bf Forecasting density-valued functional panel data}
} \fi

\bigskip

\begin{abstract}
We introduce a statistical method for modeling and forecasting functional panel data represented by multiple densities. Density functions are nonnegative and have a constrained integral and thus do not constitute a linear vector space. We implement a center log-ratio transformation to transform densities into unconstrained functions. These functions exhibit cross-sectional correlation and temporal dependence. Via a functional analysis of variance decomposition, we decompose the unconstrained functional panel data into a deterministic trend component and a time-varying residual component. To produce forecasts for the time-varying component, a functional time series forecasting method, based on the estimation of the long-run covariance, is implemented. By combining the forecasts of the time-varying residual component with the deterministic trend component, we obtain $h$-step-ahead forecast curves for multiple populations. Illustrated by age- and sex-specific life-table death counts in the United States, we apply our proposed method to generate forecasts of the life-table death counts for 51 states.

\vspace{.1in}
\end{abstract}

\newpage
\spacingset{1.6}

\section{Introduction}\label{sec:intro}

Actuaries and demographers are interested in developing models for mortality forecasting \citep[see, e.g.,][for comprehensive reviews]{BCB23}. From an actuary perspective, mortality forecasts are important inputs in pricing fixed-term or lifetime annuities and setting of reserves. They are important to manage the financial effects of mortality improvements over time on pensions and are essential to the financial viability of life insurance companies. From a demography perspective, mortality forecasts are essential to health-care and aged-care systems and are an important component of population forecasts for urban and rural planning.

For modeling age-specific mortality, three functions related to mortality are commonly studied: mortality rate, survival function, and life-table death counts. Mortality rates are bounded between 0 and 1. The survival function monotonically decreases. The life-table death counts of each age group are non-negative and sum to a radix of $10^5$ each year, which strongly resembles probability density functions (PDFs). Among these three functions, life-table death counts provide important insights into longevity risk and lifespan variability that cannot be easily quantified from either hazard or survival function. These additional constraints on the life-table death counts can be helpful to improve forecast accuracy \citep[see, e.g.,][]{Oeppen08}. For instance, in many developed countries, deaths at younger ages gradually shift toward older ages. In addition to providing an informative description of the mortality experience of a population, the life-table death counts yield readily available information on "central longevity indicators", that is, mean, median, and mode age at death \citep{CanudasRomo10}, and lifespan variability \citep{Robine01}. For many developed countries, a decrease in variability over time may be observed through the Gini coefficient of life-table ages at death  \citep[see, e.g.,][]{SHX22} or Drewnowski's index \citep[see, e.g.,][]{ABB+22}. 

Because of the two constraints, life-table death counts can be viewed as compositional data. Compositional data are defined as a random vector of $D$ compositions $[d_1, d_2, \dots,d_D]$, non-negative values in which the sum of is a given constant, typically one (portions) in PDF, 100 (\%) in weekly household expenditure \citep[see, e.g.,][]{SW17}, and $10^6$ for parts per million in geochemical compositions \citep[see, e.g.,][]{SDG+15}. The sample space of compositional data is a (nonlinear) simplex
\begin{equation}
\mathcal{S}^D = \left\{(d_1, d_2, \dots, d_D)^{\top}, \quad d_x>0, \quad \sum^D_{x=1}d_x = c\right\}, \label{simplex}
\end{equation}
where $d_x$ denotes life-table death count for age $x$, where $x=1, 2,\ldots, D$ and $D$ denotes the total number of discrete ages, $c$ is a fixed constant, and $^{\top}$ denotes vector transpose.

In compositional data analysis, there exist several transformations from the simplex to Euclidean space. Among them, the center log-ratio (clr) is the simplest and most widely used transformation because principal component analysis (PCA) is not invariant to a permutation of the compositions \citep{A1982}. For a given year $t$, the clr transformation is defined as
\begin{align}
y_{t,x}&=\ln\left\{\frac{d_{t,x}}{(\prod^D_{x=1}d_{t,x})^{1/D}}\right\}, \notag\\
&=\ln d_{t,x} - \frac{1}{D}\sum^D_{x=1}\ln d_{t,x}, \label{eq:clr}
\end{align}
where $\ln(\cdot)$ denotes natural logarithm, $\bm{y}_t = (y_{t,1}, y_{t,2}, \dots, y_{t,D})^\top$ for $t=1, 2,\dots,T$. From a time series of $\bm{Y}=(\bm{y}_1, \bm{y}_2, \dots,\bm{y}_T)$, \cite{BCO+17} applied a principal component analysis (PCA) to extract latent principal components and their associated scores. The scores can be modeled and forecast using a univariate time series forecasting method. With the forecast principal component scores, $\widehat{\bm{y}}_{T+h|T}$ can be obtained with the estimated principal components. By taking the inverse clr transformation, $\bm{d}_{T+h|T}$ can be estimated by
\[
\bm{\widehat{d}}_{T+h|T}=\frac{\exp{\left(\widehat{y}_{T+h|T,x}\right)}}{\sum^D_{x=1}\exp{\left(\widehat{y}_{T+h|T, x}\right)}}\times 10^5,
\]
where $10^5$ is the radix reported in the \cite{HMD2023}. 

Extending these works, our contribution is to develop statistical models to jointly forecast life-table death counts as a function of age at the \textit{sub-national} level. Understanding patterns in mortality across subpopulations is essential for local health policy decision-making, identifying mortality patterns of vulnerable groups, classifying heterogeneous populations into homogeneous groups, and tracking the effects of policy responses. The series at the sub-national level presents two challenges: 
\begin{inparaenum}
\item[(1)] the signal-to-noise ratio is comparably low; and
\item[(2)] the number of series may exceed the number of curves in a series. 
\end{inparaenum}
The latter leads to the study of high-dimensional functional time series (HDFTS) \citep[see, e.g.,][]{GSY19, CFJV24, CFQ+24}. 

In the context of HDFTS, \cite{ZD23} derived Gaussian and multiplier bootstrap approximations for sums of HDFTS. These approximations were utilized to construct joint simultaneous confidence bands for the mean functions and develop a hypothesis test to assess the parallel behavior of the mean functions in the panel dimension. \cite{HNT231} investigated the representation of HDFTS using a factor model, identifying crucial conditions on the eigenvalues of the covariance operator for the existence and uniqueness of the factor model. \cite{GSY19} adopted a two-stage approach, combining truncated principal component analysis and a separate scalar factor model for the resulting panels of scores. In contrast,  \cite{Tavakoli2022} introduced a functional factor model with a functional factor loading and a vector of real-valued factors, while  \cite{GQW22} considered a functional factor model with a real-valued factor loading and a functional factor. Additionally, \cite{TSY22} studied clustering problems for age-specific mortality rates as an example of HDFTS, and  \cite{LLS24} proposed hypothesis tests for detecting and identifying change points, and clustering of change points in HDFTS using an information criterion.

In the realm of estimating HDFTS models, \cite{GQ23} introduced a three-step procedure that incorporates a novel functional stability measure, the non-asymptotic properties of functional principal component analysis (FPCA), and a regularization approach for estimating autoregressive coefficients. Moreover, \cite{TLG+24} proposed dynamic weak separability to characterize the two-way dependence structure in multivariate functional time series, developing a unified framework for functional graphical models and dynamic principal component analysis. \cite{CCQ+23} presented a three-step framework for statistical learning of HDFTS with errors, incorporating autocovariance-based dimension reduction and a novel block regularized minimum distance estimation. 

In practice, it is common to find functional objects that do not reside in a linear Hilbert space. This is exemplified by various scenarios, such as analyzing time series consisting of PDFs and cumulative density functions (CDFs). In the context of PDFs, the works of \cite{PLC2019} and \cite{SHX22} have studied the age distribution of death counts. In particular, \cite{SHX22} have proposed a forecasting method that utilizes the location and shape measures of density functions. Moreover, in nonstationary time series, \cite{CKP16} have considered modeling state densities, while \cite{HZ18} have explored the dynamics of financial return densities. In terms of CDFs, \cite{Condino23} have examined a sequence of Lorenz curves derived from a regional household income and wealth survey in Italy. \cite{QMA24} proposed a functional ridge-regression-based estimation method that estimates CDFs accurately. We aim to introduce a forecasting method for HDFTS where the functional objects are PDFs.

When the functional objects are nonlinear, it is common to transform the functional objects into a linear space. A popular space is the Bayes Hilbert space studied in \cite{HMT+16}. \cite{HMT+16} introduced the simplicial FPCA of densities. Simplicial FPCA is based on the geometry of the Bayes space of functional compositions. It utilizes the clr transformation to map the densities to a suitable Hilbert space, enabling the application of standard functional data analytic techniques. \cite{SB19} focused on modeling nonstationary time series of PDFs. They illustrated the isomorphism between a cointegrated Bayes Hilbert space and a cointegrated linear process in a Hilbert space of centered square-integrable real functions.

A unique feature associated with the HDFTS is that they can be cross-sectionally correlated and temporally dependent. We deploy a two-way functional analysis of variance (ANOVA), as used by \cite{CFJV24} to decompose HDFTS into a deterministic trend component and a residual component that varies over time. These components can be estimated by functional median polish decomposition (FMP-ANOVA) and functional analysis of variance based on means (FM-ANOVA). Both decompositions compute the functional grand effect and functional main factor effects in an additive model without factor interaction. The FM-ANOVA has been extensively studied by \cite{Zhang14}. The FMP-ANOVA was proposed by \cite{SG12} as an extension of the univariate median polish of \cite{tukey1977}. It is a robust statistical technique for studying the effects of factors on the response since it replaces the mean with the median. Our proposed functional time series forecasting method for densities is then compared with existing methods, such as the factor models from \cite{GSY19} and \cite{Tavakoli2022}, and the maximum entropy mortality model proposed by \cite{PLC2019}.

The remainder of this paper is structured as follows. In Section~\ref{sec:empirical}, we present the United States (US) sub-national mortality rates across 51 states. In Section~\ref{sec:methods}, we apply functional ANOVA decomposition to extract a deterministic trend component and a time-varying residual component. For modeling and forecasting the residual component, we introduce a functional time series forecasting method based on FPCA for producing forecasts. By combining the time-varying component forecasts with the deterministic component, we obtain $h$-step-ahead forecast curves for multiple populations. With the inverse clr transformation, we transform our forecast curves back to the simplex. In doing so, we obtain forecast life-table death counts. We evaluate and compare point forecast accuracies in Section~\ref{sec:PF_comp}. Section~\ref{sec:conclusion} concludes and offers some ideas on how the methodology presented can be further extended.

\section{Age distribution of life-table death counts in the United States}\label{sec:empirical}

We consider US age- and sex-specific life-table death counts from 1959 to 2020, obtained from the \cite{USMD23}. The database documents a historical set of complete state-level life tables designed to foster research on geographic variations in mortality across the US and monitor trends in health inequalities. This data set includes complete and abridged life tables by sex for each of the nine Census Divisions, four Census Regions, 50 States of the US, as well as the District of Columbia, with mortality up to age 110. 

The life-table radix is fixed at $10^5$ at age 0 for each year. For the life-table death counts, there are 111 ages, and these are age $0, 1, \dots, 109, 110+$. Due to rounding, there may be zero counts, especially for older ages at some years. To rectify the problem, one can also use the probability of dying (i.e., $q_x$) and the life-table radix to recalculate the estimated death counts with decimal places. The smoother estimates are more detailed than those reported in the \cite{USMD23}. 
\begin{figure}[!htb]
\centering
\subfloat[Rhode Island data series between 1959 and 2020]
{\includegraphics[width=8.6cm]{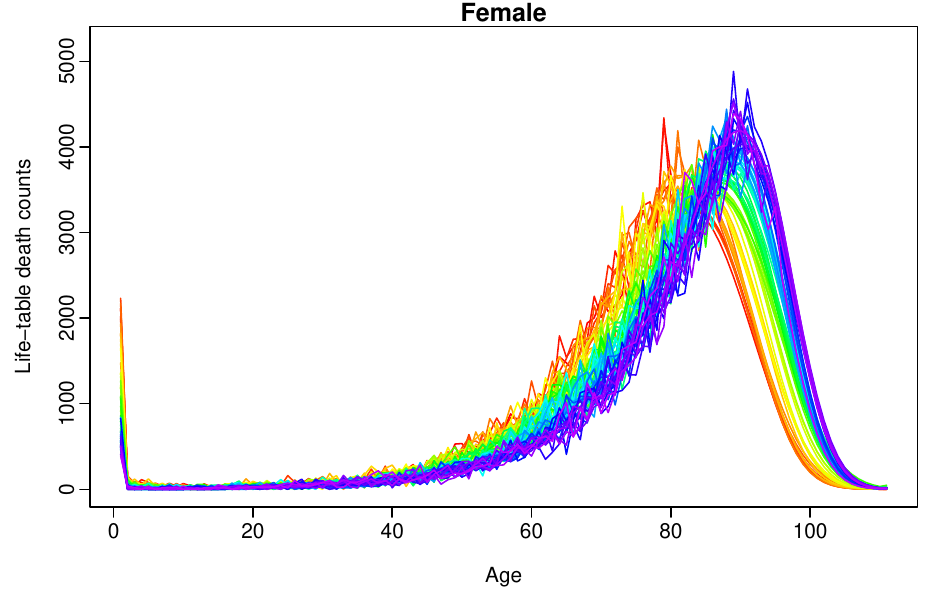}\label{fig:1a}}
\quad
\subfloat[Rhode Island data series between 1959 and 2020]
{\includegraphics[width=8.6cm]{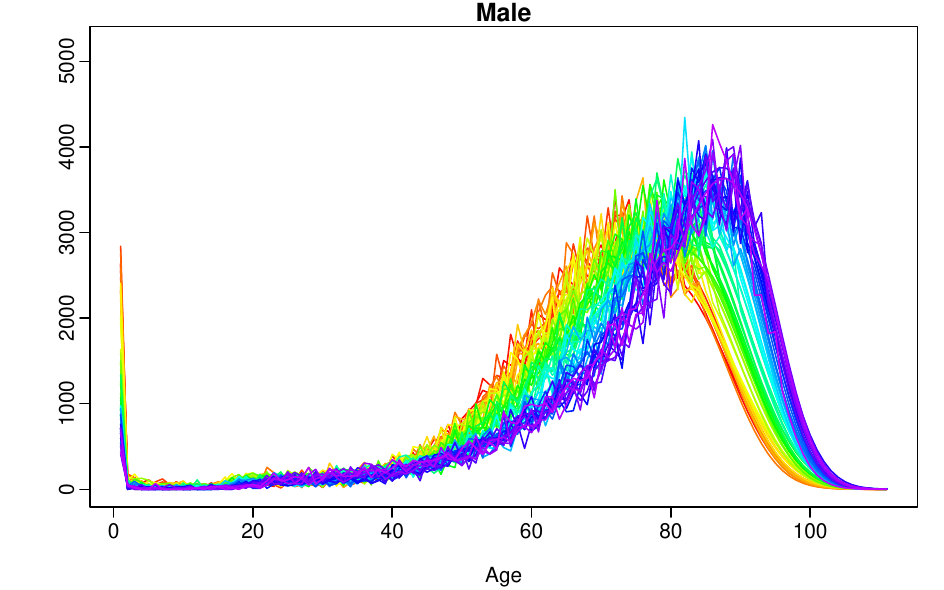}\label{fig:1b}}
\caption{\small{Rainbow plots of age-specific life-table death count from 1959 to 2020 in a single-year group in Rhode Island,US. The oldest years are shown in red, with the most recent in violet. Curves are ordered chronologically according to the colors of the rainbow.}}
\end{figure}

To understand the primary features of the data, Figures~\ref{fig:1a} and~\ref{fig:1b} present rainbow plots of the female and male age-specific life-table death counts in Rhode Island in a single-year group, respectively. The time ordering of the curves follows the colors of the rainbow, where data from the distant past are shown in red and more recent data are shown in purple. Both figures demonstrate a decreasing trend in infant mortality and a negatively skewed distribution for the life-table death counts, where the peaks shift to older ages for both females and males. The shift of the distribution symbolizes longevity risk, which can be a major issue for insurers and government pensions, especially in the selling and risk management of annuities \citep[see][for discussion]{BDV02, CanudasRomo10}. 

\begin{figure}[!htb]
\centering
{\includegraphics[width=8.65cm]{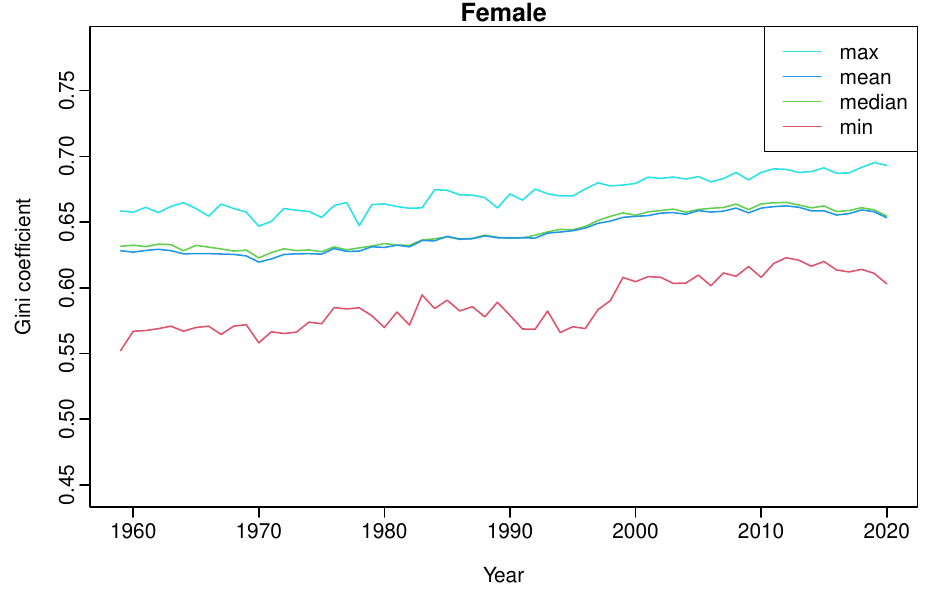}\label{fig:2a}}
\quad
{\includegraphics[width=8.65cm]{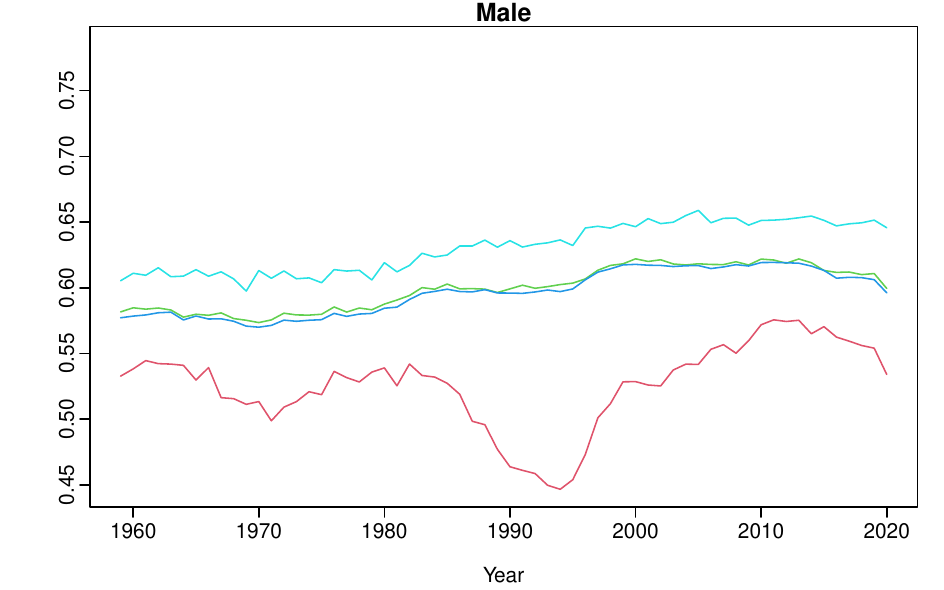}\label{fig:2b}}
\caption{\small{Summary statistics (maximum, mean, median, and minimum) for the Gini coefficients in the US aggregated across the 51 states.}\label{fig:2}}
\end{figure}

To measure the spread of the distribution, we consider the Gini coefficient. It is a measure that quantifies the level of inequality in age at death within a life table population. A lower Gini coefficient value indicates less age variability at death, suggesting a greater level of equality \citep{Basellini2020,ABB+22}. In Figure~\ref{fig:2}, we present the Gini coefficient for the US for the female population (left) and male population (right). We can observe from both figures that, on average, there has been an increasing trend in the Gini coefficient over the years. The observed increasing tendency can be explained by a continuous decrease in the ratio between the resulting life expectancy at birth when doubling the hazard at all ages and the overall life expectancy at birth \citep{ABB+22}. 

\section{Methodology}\label{sec:methods}

The life-table death counts are discrete and densely observed. By adopting the linear interpolation, we convert them into a continuous function on the same function support $u\in (x_1, x_D)$, where $x_1$ and $x_D$ represent the first and last age groups, respectively. 


\subsection{Center log-ratio transformation for density-valued functional time series}\label{clr}

Let $d_{t,s}^g(u)$ be the age-specific life-table death count for age~$u$, state~$s$, gender~$g$ at year~$t$. For a given year $t$, compositional data are defined as a random vector of non-negative components, and the sum of which is a specified constant. Between the non-negativity and summability constraints, the sample space of functional compositional data is a simplex of the form in \eqref{simplex}.

When age $u\in [x_1, x_D]$ is treated as a continuum, the clr transformation in~\eqref{eq:clr} can be written as 
\begin{equation*}
\text{clr}[d^g_{t,s}(u)] := \Y_{t,s}^g(u) =  \ln d^g_{t,s}(u) - \frac{1}{\eta}\int^{x_D}_{x_1} \ln d^g_{t,s}(u)du, \label{eq:clr_transf}
\end{equation*}
where $\eta=x_D-x_1$ denotes the length of the age interval, and $\frac{1}{\eta}\int^{x_D}_{x_1} \ln d^g_{t,s}(u)du$ is the geometric mean. Let $\Y_{t,s}^g(u)$ be the transformed data for age $u$, state $s$, gender $g$ at year $t$.  When considering life-table death counts as density curves, \cite{SH20} and \cite{SHX22} developed a functional principal component regression to forecast a time series of density curves in the constrained space. Via the inverse clr transformation, the forecast life-table death counts can be obtained as
\[
\widehat{d}^g_{T+h|T,s}(u) = \frac{\exp[\widehat{\Y}_{T+h|T,s}^g(u)]}{\int^{x_D}_{x_1} \exp[\widehat{\Y}_{T+h|T,s}^g(u)] du}\times 10^5. 
\]

\subsection{Two-way functional ANOVA model}\label{FMP}

By treating age $u$ as a continuum, a two-way functional ANOVA decomposition can be expressed as: 
\begin{equation}
\Y_{t,s}^{g}(u)=\underbrace{\mu(u)+\alpha_s(u)+\beta^g(u)}_{\text{Deterministic:=}~\mathcal{D}_{s}^{g}(u)}+\underbrace{\X^{g}_{t,s}(u)}_{\text{time-varying}}, \qquad u\in \mathcal{I}\subset R. \label{eq:2}
\end{equation}
where $\mu(u)$ denotes the functional grand effect, $\alpha_{s}(u)$ denotes the functional row effect, and, $\beta^{g}(u)$ denotes the functional column effect.  
For each state $s$ and gender $g$, we considered replicates for the years $t$ with time horizon, for $t=1, 2,\ldots, T$, $s=1, 2,\ldots,n_s$, and $g=\{\text{Female}(\text{F}), \text{Male}(\text{M})\}$. $\bm{\X}_{s}^{g}(u)=[\X_{1,s}^{g}(u),\dots,\X_{T,s}^{g}(u)]$ denotes the residual component that varies over time for the state~$s$ and gender~$g$. Such a decomposition is not an approximation and can exactly reconstruct the original data.  

The model described in~\eqref{eq:2}, can be estimated in two ways: the FM-ANOVA approach \citep{Zhang14, RS06} or FMP-ANOVA approach \citep{SG12}. The former extracts the deterministic component into the following terms:
\begin{align*}
\widehat{\mu}(u)&=\frac{1}{n_s\times 2 \times T}\sum_{s=1}^{n_s}\sum_{g=1}^{2}\sum_{t=1}^{T} \Y_{t,s}^{g}(u)\\
\widehat{\alpha}_s(u)&= \frac{1}{2\times T}\sum_{g=1}^{2}\sum_{t=1}^{T} \Y_{t,s}^{g}(u)-\widehat{\mu}(u)\\
\widehat{\beta}^{g}(u)&= \frac{1}{n_s\times T}\sum_{s=1}^{n_s}\sum_{t=1}^{T} \Y_{t,s}^{g}(u)-\widehat{\mu}(u).
\end{align*}
There exist some identifiability constraints, so that for all $u \in \mathcal{I}$, $\sum_{s=1}^{n_s}\alpha_s(u)=\sum_{g=1}^{2}\beta^{g}(u)=0$, and $\sum_{s=1}^{n_s} \X^{g}_{t,s}(u)=\sum_{g=1}^{2}\X^{g}_{t,s}(u)=0$ for all $t$ \citep[][Chapter 13]{RS06}. The FMP-ANOVA decomposition satisfies that  $\forall u \in \mathcal{I}$, $\text{median}_s\{ \alpha_s(u)\}=0$, $\text{median}_g\{ \beta^{g}(u)\}=0$, $\text{median}_s\{ \X^{g}_{t,s}(u)\}=\text{median}_g\{ \X^{g}_{t,s}(u)\}=0$ for all $t$ \citep{SG12}.

\subsection{Point forecasts based on functional analysis of variance}\label{PF_FM/FMP}

For the functional residual process obtained from either the FMP-ANOVA or FM-ANOVA, we consider a functional time series forecasting method based on the FPCA of \cite{Shang19}. The proposed FPCA relies on an accurate estimate of the long-run covariance function, for which we consider a kernel sandwich estimator with plug-in bandwidth \citep[see, e.g.,][]{Horváth2016, Rice2017}. 

For a given state $s$ and gender $g$, denote $\X_{t,s}^g(u)$ as a stationary ergodic functional time series exhibiting stationarity and ergodicity. In essence, the statistical features of a stochastic process will not vary over time, and they can be obtained from a single, sufficiently long sample of the process. For such a random process, the long-run covariance function can be defined as
\begin{align*}
C_{s}^{g}(u,v)&=\sum_{l=-\infty}^{\infty} \gamma_{l,s}^{g}(u,v)\\
&=\sum_{l=-\infty}^{\infty} \text{cov}\Big[\X_{0,s}^g(u),\X_{l,s}^g(v)\Big],
\end{align*}
where $u, v \in \mathcal{I}$ and $l$ denote a lag variable. 

While the long-run covariance can be expressed as a bi-infinite summation, its estimation is not trivial. For a finite sample, a natural estimator of $C(u,v)$ is
\begin{equation}
\widehat{C}_{T,s}^{g}(u,v)=\frac{1}{T}\sum_{|l|=0}^{|l| \leq T} (T-|l|)\widehat{\gamma}_{l,s}^{g}(u,v),\label{hat_cov}
\end{equation}
where $\overline{\X}_{s}^{g}=\frac{1}{T}\sum^{T}_{t=1}\X_{t,s}^g(u)$ and
\begin{equation*}
\widehat{\gamma}_{l,s}^{g}(u,v)=\begin{cases}
\frac{1}{T}\sum_{t=1}^{T-l}\big[\X_{t,s}^g(u)-\overline{\X}_s^g(u)\big]\big[\X_{t+l,s}^{g}(v)-\overline{\X}_s^g(v)\big] & \text{ if } l \geq 0; \\ 
\frac{1}{T}\sum_{t=1-l}^{T}\big[\X_{t,s}^g(u)-\overline{\X}_s^g(v)\big]\big[\X_{t+l,s}^{g}(v)-\overline{\X}_s^g(v)\big] & \text{ if } l<0. 
\end{cases}
\end{equation*}

The long-run covariance function in~\eqref{hat_cov} can be seen as a sum of autocovariance functions with decreasing weights. It is common in practice to determine the optimal lag value of $l$ to balance the trade-off between squared bias and variance. In \cite{Li2020}, $l$ is the minimum between sample size $T$ and the number of discretized points in a function. Other approaches use the kernel sandwich estimator as in \cite{Horváth2016}
\begin{equation*}
\widehat{\widehat{C}}_{T,b,s}^{g}(u,v)= \sum_{l=-\infty}^{\infty} W_q\left( \frac{l}{b}\right)\widehat{\gamma}_{l,s}^{g}(u,v),
\end{equation*}
where $b$ is the bandwidth parameter, and $W_q(\cdot)$ is a symmetric weight function with bounded support of order $q$. \cite{Rice2017} propose a plug-in algorithm for obtaining the optimal bandwidth parameter to minimize the asymptotic mean-squared normed error between the estimated and theoretical long-run covariance functions.

Via Mercer's lemma \citep{Mercer}, the estimated long-run covariance function $\widehat{\widehat{C}}_{T,b}(u,v)$ can be approximated by 
\begin{equation*}
\widehat{\widehat{C}}_{T,b,s}^{g}(u,v)=\sum_{k=1}^{\infty} \theta_{k} \phi_{k,s}^g(u)\phi_{k,s}^g(v),
\end{equation*}
where $\theta_1 \geq \theta_2 \geq \ldots \geq 0$ are the eigenvalues of $\widehat{\widehat{C}}_{T,b,s}(u,v)$, and $[\phi_{1,s}^{g}(u),\phi_{2,s}^{g}(u),\ldots]$ are the orthonormal functional principal components. We can project a functional time series onto a set of orthonormal functional principal components via the inner product in the corresponding Hilbert space. This leads to the Karhunen-Lo\`{e}ve expansion of  the realization of a stochastic process,
\begin{align*}
\X_{t,s}^g(u)&=\overline{\X}_s^g(u)+\sum_{k=1}^{\infty} \gamma_{k,t,s}^g\phi^{g}_{k,s}(u)\\
&=\overline{\X}_s^g(u)+\sum_{k=1}^{K} \gamma_{k,t,s}^g\phi^{g}_{k,s}(u)+\epsilon_{t,s}^g(u),
\end{align*}
where $\gamma^{g}_{k,t,s}=\bigl \langle \X_{t,s}^g(u)-\overline{\X}_s^g(u),\phi^{g}_{k,s}(u) \bigr \rangle$, denotes the $k^{th}$ set of principal component scores for year~$t$, sex $s$ and geographical region $g$, and $\langle \cdot, \cdot \rangle$ denotes inner product. In addition, $\epsilon_{t,s}^g(u)$ denotes the error term with zero mean and finite variance, and $K$ denotes the number of retained components.

There are various ways to determine the value of $K$: 
\begin{inparaenum}
\item[(1)] scree plots or the fraction of variance explained by the first few functional principal components \citep{Chiou12};
\item[(2)] pseudo-versions of Akaike information criterion and Bayesian information criterion \citep{YMW05};
\item[(3)] predictive cross validation leaving out one or more curves \citep{RS06};
\item[(4)] bootstrap methods \citep{HV06}; and
\item[(5)] eigenvalue ratio criterion \citep{AH13, LRS21}.
\end{inparaenum}
Here, we select $K$ via the eigenvalue ratio criterion (EVR) of \cite{LRS21}. The value of $K$ is determined as the integer minimizing ratio of two adjacent empirical eigenvalues given by
\begin{equation}
\widehat{K}=\argmin_{1 \leq \kappa \leq \kappa_{\text{max}}} \Biggl\{ \frac{\widehat{\theta}_{\kappa+1}}{\widehat{\theta}_{\kappa}} \times \mathbbm{1} \left( \frac{\widehat{\theta}_{\kappa+1}}{\widehat{\theta}_{\kappa}} \geq \delta \right) + \mathbbm{1}  \left( \frac{\widehat{\theta}_{\kappa+1}}{\widehat{\theta}_{\kappa}} \leq \delta \right)\Biggr\}, \label{eq:eigen_ratio}
\end{equation}
where $\delta$ is a pre-specified small positive number, set as $\delta=1/\ln(\text{max}\{\widehat{\theta}_1,T\})$, and $\mathbbm{1}(\cdot)$ is the binary indicator function. $\kappa_{\text{max}}$ is a pre-specified positive integer, we choose $\kappa_{\text{max}}$, as follows
\begin{equation*}
\kappa_{\text{max}}=\Big|\Big\{ \kappa|\widehat{\theta}_{\kappa} \geq \frac{1}{T}\sum_{\kappa=1}^T \widehat{\theta}_{\kappa}, \quad \kappa \geq 1 \Big\}\Big|,
\end{equation*}
where $|\cdot|$ denotes the cardinality of the set.

Because the female and male populations within a state can exhibit a strong dependence, we utilize a multivariate FPCA previously studied in \cite{SK22}. For a given state~$s$, collectively modeling multiple populations requires truncating $K$ numbers of functional principal components of joint time series 
\[
\bm{\X}_{t,s}(u) \approx \bm{\Phi}_s(u)\bm{\Gamma}_{t,s},
\]
where $\bm{\X}_{t,s}(u) = [\X_{t,s}^{\text{F}}(u),\X_{t,s}^{\text{M}}(u)]^{\top}$, superscripts \textsuperscript{F} and \textsuperscript{M} denote the female and male populations in a given state, and
\begin{equation*}
   \bm{\Gamma}_{t,s}=\left[\gamma^{\text{F}}_{1,t,s}, \ldots, \gamma^{\text{F}}_{K,t,s},\gamma^{\text{M}}_{1,t,s},\ldots, \gamma^{\text{M}}_{K,t,s}\right]^{\top},
\end{equation*}
is a $(2 \times K) \times 1$ vector of stacked principal component scores. $\bm{\Phi}_s(u)$ is a $2 \times (2 \times K)$ matrix that contains the associated orthogonal basis functions. By conditioning on $\bm{\Phi}_s(u)$, we obtain the $h$-step-ahead point forecasts as follows
\begin{align*}
\widehat{\bm{\X}}_{T+h|T,s}(u)&=\mathds{E}\Big[\bm{\X}_{T+h,s}(u)\big|\bm{\X}_{1,s}(u),\ldots,\bm{\X}_{T,s}(u);\bm{\Phi}_s(u)\Big]\\
&=\bm{\overline{\X}}_s(u)+\bm{\Phi}_s(u)\widehat{\bm{\Gamma}}_{T+h|T,s},
\end{align*}
where $\bm{\overline{\X}}_s(u)=[\overline{\X}_s^{\text{F}}(u), \overline{\X}_s^{\text{M}}(u)]$ represents the joint sample mean function. 

A univariate time series method can be implemented to produce the forecasts of the principal component scores. The most commonly used methods are the exponential smoothing and autoregressive integrated moving average (ARIMA) models. For obtaining the forecast principal component score $\widehat{\bm{\Gamma}}_{T+h|T,s}$, we apply the optimal ARIMA model selected by \cite{Hyndman2008}. Because the age-specific life-table death counts are observed yearly, the ARIMA takes the generic form of
\begin{equation}
\left(1-\tau_1\mathcal{B}-\cdots-\tau_{\textit{p}}\mathcal{B}^{\textit{p}}\right)(1-\mathcal{B})^d\gamma_{\kappa}= \psi+(1+\nu_1\mathcal{B}+\cdots+\nu_{\textit{q}}\mathcal{B}^{\textit{q}})\omega_{\kappa},
\label{eq:ARIMA}
\end{equation}
where $\psi$ represents the intercept, $(\tau_1,\ldots,\tau_{\textit{p}})$ denote the coefficients associated with the autoregressive component, $\gamma_{\kappa}$ represents the principal component score, $(\nu_1,\ldots,\nu_{\textit{q}})$ denote the coefficients associated with the moving average component, $\mathcal{B}$ denotes the backshift operator, $(1-B)$ denotes the differencing operator, $d$ represents the degree of integration, and $\omega_{\kappa}$ represents a white-noise error term. The automatic selection method of \cite{Hyndman2008} chooses the optimal autoregressive order $\textit{p}$, moving average order $\textit{q}$, and difference order $d$. The value of $d$ is chosen using the Kwiatkowski-Phillips-Schmudt-Shin (KPSS) unit root tests \citep{Kwiatkowski1992}. 

Once the forecasted functional residuals are obtained, we add the deterministic component from the FMP-ANOVA decomposition. As this is not time-varying,  the overall $h$-step-ahead point forecast is defined as
\begin{equation*}
\widehat{\Y}_{T+h|T,s}^{g}(u)=\mathcal{D}_{s}^g(u) + \widehat{\X}_{T+h|T, s}^g(u),    
\end{equation*}
where $\mathcal{D}_{s}^g(u)$ is defined in~\eqref{eq:2}. Then, we transform back to the original scale through the inverse clr transformation in Section~\ref{eq:clr_transf}.

\section{Forecast accuracy evaluation of life-table death counts}\label{sec:PF_comp}


We consider a rolling-window scheme to assess the point forecast as described in \citet[][Chapter 9]{zivot2006}. The procedure is carried out as follows.
\begin{enumerate}
\item[1)] The transformed data are decomposed, through the two-way FMP-ANOVA and FM-ANOVA in Section~\ref{FMP}, into deterministic and time-varying components. The two factors are the state~$s$ and gender~$g$ (males and females). The functional residual curves $[\X_{1,s}^{g}(u),\dots,\X_{T,s}^g(u)]$ are those obtained after removing all deterministic components.
\item[2)]  We perform an $h$-step-ahead point forecast of the time-varying component. Then, we add the deterministic components to obtain the point forecast of the future curves.
\item[3)] To compute each of the $h$-step-ahead point forecasts, for $h=1, 2,\ldots, H$, we proceed as follows: we consider a rolling window as a training set of size $T$ and produce a $(T+h)$-step-ahead point forecast. 
\item[4)] The process iterates from $T+1$ to $T+H$, where the training data set with equal sample size rolls one-step-ahead each time.
\item[5)] With the holdout densities, we compute the point and interval forecast errors.
\end{enumerate}

Since the sub-national age-specific life-table death counts can be considered a PDF, we consider some density evaluation measures to evaluate the point forecast accuracy. These metrics include the discrete version of the Kullback-Liebler divergence \citep{KLD1951} and the Jensen-Shannon divergence \citep{Shannon1948}. The Kullback-Liebler divergence is intended to measure the loss of information when we choose an approximation. For two PDFs denoted by $d^{g}_{T+h, s}(u)$ and $\widehat{d}_{T+h|T,s}^{g}(u)$, the discrete version of the Kullback-Liebler divergence is defined as
\begin{equation*}
\text{KLD}^g_s(h) =\text{D}_{\text{KL}}\left[d^{g}_{T+h, s}(u_i)||\widehat{d}_{T+h|T,s}^{g}(u_i)\right] + \text{D}_{\text{KL}} \left[\widehat{d}_{T+h|T,s}^{g}(u_i)||d^{g}_{T+h, s}(u_i)\right],
\end{equation*}
which is symmetric and non-negative, and $i$ represents a discretized age. An alternative is given by the Jensen-Shannon divergence defined by
\begin{equation*}
\text{JSD}^g_s(h) =  \frac{1}{2}\text{D}_{\text{KL}}\left[d^{g}_{T+h, s}(u_i)||\delta^g_{T+h, s}(u_i) \right] +\frac{1}{2}\text{D}_{\text{KL}}\left[\widehat{d}_{T+h|T,s}^{g}(u_i)||\delta^g_{T+h, s}(u_i) \right],
\end{equation*}
where $\delta^g_{T+h, s}(u_i)$ measures a common quantity between $d^{g}_{T+h, s}(u_i)$ and $\widehat{d}_{T+h|T,s}^{g}(u_i)$  \citep{Fuglede2004}. We consider geometric mean given by  $\delta_{T+h, s}^g(u_i)=\sqrt{d^g_{T+h, s}(u_i)\widehat{d}_{T+h|T,s}^{g}(u_i)}$. 

We report the averaged KLD and JSD across $n_s$ number of states. That is, 
\begin{align*}
\text{KLD}^g(h) = \ & \frac{1}{n_s\times p}\sum_{s=1}^{n_s}\sum_{i=1}^{p} d^{g}_{T+h, s}(u_i) 
                \cdot [\ln{d^{g}_{T+h, s}(u_i)}-\ln{\widehat{d}_{T+h|T,s}^{g}(u_i)}]+\\
                  & \frac{1}{n_s\times p}\sum_{s=1}^{n_s} \sum_{i=1}^{p} \widehat{d}_{T+h|T,s}^{g}(u_i)  \cdot [\ln{\widehat{d}_{T+h|T,s}^{g}(u_i)}-\ln{d_{T+h,s}^{g}(u_i)}].
\end{align*}
The average across the forecast horizons for both KLD and JSD can be computed as follows,
\begin{align*}
\overline{\text{KLD}}^g&=\frac{1}{H}\sum_{h=1}^{H} \text{KLD}^g(h)\\
\overline{\text{JSD}}^g&=\frac{1}{H}\sum_{h=1}^{H} \text{JSD}^g(h).
\end{align*}

\subsection{Point forecast comparison}\label{Comp:PF}

The results for the point forecast accuracy for the US, using the EVR criterion described in~\eqref{eq:eigen_ratio}, are provided in Table~\ref{tab:1}. We present the KLD and $\text{JSD}$ across the forecast horizon averaged across states and gender. We incorporate the accuracy mean over the forecast horizon. In Appendix~\ref{A1:K_6}, we conduct additional sensitivity analysis by fixing the number of functional principal components to $K=6$ as advocated by \cite{HBY13}.
\begin{center}
\tabcolsep 0.228in
\begin{longtable}{@{}lllllllll@{}}
\caption{\small Point forecast accuracy, as measured by the KLD and $\text{JSD}$, for the US between the FMP-ANOVA and FM-ANOVA. All values are multiplied by a factor of $100$.} \label{tab:1}\\
\toprule
& \multicolumn{4}{c}{\textbf{FMP-ANOVA}} & \multicolumn{4}{c}{\textbf{FM-ANOVA}} \\
& \multicolumn{2}{c}{\textbf{Female}} & \multicolumn{2}{c}{\textbf{Male}} & \multicolumn{2}{c}{\textbf{Female}} & \multicolumn{2}{c}{\textbf{Male}}  \\ \cmidrule{2-9}
$h$ & \textbf{KLD}       & \textbf{JSD}    & \textbf{KLD}         & \textbf{JSD} & \textbf{KLD}       & \textbf{JSD}    & \textbf{KLD}         & \textbf{JSD}   \\ \midrule
\endfirsthead
\toprule
& \multicolumn{4}{c}{\textbf{FMP-ANOVA}} & \multicolumn{4}{c}{\textbf{FM-ANOVA}} \\
& \multicolumn{2}{c}{\textbf{Female}} & \multicolumn{2}{c}{\textbf{Male}} & \multicolumn{2}{c}{\textbf{Female}} & \multicolumn{2}{c}{\textbf{Male}}  \\ \cmidrule{2-9}
$h$ & \textbf{KLD}       & \textbf{JSD$^g$}    & \textbf{KLD}         & \textbf{JSD$^g$} & \textbf{KLD}       & \textbf{JSD$^g$}    & \textbf{KLD}         & \textbf{JSD$^g$}   \\ \midrule
\endhead
\midrule
\multicolumn{9}{r}{{Continued on next page}} \\ 
\endfoot
\endlastfoot
  1 & 1.83 & 0.52 & 2.17 & 0.59 & 1.17 & 0.33 & 1.36 & 0.36 \\ 
  2 & 1.83 & 0.52 & 2.24 & 0.61 & 1.22 & 0.34 & 1.41 & 0.38 \\ 
  3 & 1.93 & 0.53 & 2.35 & 0.64 & 1.31 & 0.37 & 1.51 & 0.40 \\ 
  4 & 2.11 & 0.58 & 2.54 & 0.68 & 1.47 & 0.40 & 1.66 & 0.44 \\ 
  5 & 2.34 & 0.64 & 2.75 & 0.73 & 1.67 & 0.46 & 1.85 & 0.49 \\ 
  6 & 2.64 & 0.72 & 3.04 & 0.80 & 1.92 & 0.52 & 2.11 & 0.55 \\ 
  7 & 2.98 & 0.82 & 3.33 & 0.88 & 2.19 & 0.59 & 2.35 & 0.61 \\ 
  8 & 3.33 & 0.90 & 3.71 & 0.98 & 2.53 & 0.67 & 2.68 & 0.70 \\ 
  9 & 4.58 & 1.37 & 4.14 & 1.09 & 3.18 & 0.86 & 3.08 & 0.80 \\ 
  10& 6.04 & 1.60 & 5.56 & 1.45 & 4.84 & 1.25 & 4.47 & 1.14 \\  \midrule
Mean &   2.66 & 0.75 & 3.18 & 0.85 & 1.95 & 0.53 & 2.28 & 0.59   \\
\bottomrule          
\end{longtable}
\end{center}

\vspace{-.4in}

From the results in Table~\ref{tab:1}, the proposed FM-ANOVA method generally outperforms the FMP-ANOVA method. The FM-ANOVA method consistently achieves lower values for KLD and $\text{JSD}$. Additionally, the accuracy of point forecasts shows an increasing trend as the forecast horizon extends. When comparing the two proposed methods, the forecast errors for the female population are generally lower than that for the male population.

\subsection{Competitive HDFTS forecasting methods}\label{Comp_meth}

We discuss alternative methods from the recent HDFTS literature. For the methods introduced by \cite{GSY19} and \cite{Tavakoli2022}, we incorporate the clr transformation to ensure that the density functions belong to the same functional space during modeling and forecasting. However, the method proposed by \cite{PLC2019} does not require transformation, as it is designed to work directly with densities. In Appendix~\ref{A2:wo_clr}, we also included the results of the two alternative methods without the clr transformation.

\subsubsection{Two-stage functional principal component analysis}

\cite{GSY19} adopted a two-stage approach combining truncated principal component analysis and a separate scalar factor model for the resulting panels of scores. Their method consists of three steps.
\begin{enumerate}
\item[1)] Dynamic FPCA is performed separately on each set of functional time series, resulting in $N$ sets of principal component scores of low dimension $p_0$.
\item[2)] The first functional principal component scores from each of $N$ sets of functional time series are combined into an $N\times 1$ vector. We fit factor models to the FPC scores to further reduce the dimension into an $r\times 1$ vector, where $r<< N$. The same procedure is implemented for the second, third, and so on until the $p_0$\textsuperscript{th} FPC scores. The vector of $N$ functional time series is reduced to an $r\times p_0$ matrix.
\item[3)] A scalar time series model can be fitted to each factor and forecasts are produced. The forecast factors can be used to construct forecast functions.
\end{enumerate}

\subsubsection{Functional factor model}

For a given gender $g$, \cite{Tavakoli2022} introduce the following functional factor model:
\begin{equation*}
\X_{st}(u) = \sum^{k_{\dagger}}_{k=1}B_{sk}^{\dagger}(u)F_{tk}^{\dagger}+\varepsilon_{st}(u),
\end{equation*}
where $B_{sk}^{\dagger}(\cdot)$ is the $k$\textsuperscript{th} functional factor loading, $\bm{F}_t^{\dagger} = (F_{t1}^{\dagger},\dots,F_{tk_{\dagger}}^{\dagger})^{\top}$ is a vector of real-valued factors, $\varepsilon_{st}(u)$ denotes the model residuals, and $k_{+}$ denotes the number of retained functional principal components.

\subsubsection{Maximum entropy mortality model}

In the demographic literature, several attempts exist to model the age distribution of death counts. \cite{PLC2019} propose matching finite (typically, the first four) statistical moments of the age distribution of death counts. The forecast life-table death counts for a population can be determined by extrapolating these moments. The extrapolation is achieved through multivariate time series models, such as multivariate random walk with drift. With the forecast moments, the forecast age distribution of death counts is then obtained.

Table~\ref{tab:1_new} provides an evaluation and comparison of the point forecast accuracy for three competitive methods. In the case of the methods proposed by \cite{Tavakoli2022} and \cite{GSY19}, the clr transformation is applied, as these methods were not originally designed for density-valued functions. To emphasize the significance of the clr transformation, we also present the point forecast results of these two methods without the transformation in Appendix~\ref{A2:wo_clr}. 

\begin{center}
\tabcolsep 0.088in
\begin{longtable}{@{}lllllllllllll@{}}
\caption{\small Point forecast accuracy, as measured by the KLD and $\text{JSD}$, for the US for alternative methods TNH refers to the method of \cite{Tavakoli2022}, GSY refers to the method of \cite{GSY19}, and PLC refers to the method of \cite{PLC2019}. All values are multiplied by a factor of $100$.} \label{tab:1_new}\\
\toprule
& \multicolumn{4}{c}{\textbf{TNH23}} & \multicolumn{4}{c}{\textbf{GSY19}} &\multicolumn{4}{c}{\textbf{PLC19}} \\
& \multicolumn{2}{c}{\textbf{Female}} & \multicolumn{2}{c}{\textbf{Male}} & \multicolumn{2}{c}{\textbf{Female}} & \multicolumn{2}{c}{\textbf{Male}} & \multicolumn{2}{c}{\textbf{Female}} & \multicolumn{2}{c}{\textbf{Male}}  \\ \cmidrule{2-13}
$h$ & \textbf{KLD} & \textbf{JSD} & \textbf{KLD} & \textbf{JSD} & \textbf{KLD} & \textbf{JSD}    & \textbf{KLD} & \textbf{JSD} & \textbf{KLD} & \textbf{JSD} & \textbf{KLD} & \textbf{JSD}   \\ \midrule
\endfirsthead
\toprule
& \multicolumn{4}{c}{\textbf{TNH23}} & \multicolumn{4}{c}{\textbf{GSY19}} &\multicolumn{4}{c}{\textbf{PLC19}} \\
& \multicolumn{2}{c}{\textbf{Female}} & \multicolumn{2}{c}{\textbf{Male}} & \multicolumn{2}{c}{\textbf{Female}} & \multicolumn{2}{c}{\textbf{Male}} & \multicolumn{2}{c}{\textbf{Female}} & \multicolumn{2}{c}{\textbf{Male}} \\ \cmidrule{2-13}
$h$ & \textbf{KLD}       & \textbf{JSD$^g$}    & \textbf{KLD}         & \textbf{JSD$^g$} & \textbf{KLD}       & \textbf{JSD$^g$}    & \textbf{KLD}         & \textbf{JSD$^g$} & \textbf{KLD}       & \textbf{JSD$^g$}    & \textbf{KLD}         & \textbf{JSD$^g$}   \\ \midrule
\endhead
\midrule
\multicolumn{13}{r}{{Continued on next page}} \\ 
\endfoot
\endlastfoot
  1 & 1.00 & 0.28 & 0.93 & 0.25 & 6.84 & 1.77 & 7.58 & 1.90 & 2.38 & 0.52 & 1.89 & 0.66 \\
  2 & 1.04 & 0.29 & 1.01 & 0.27 & 8.97 & 2.41 & 9.71 & 2.39 & 2.38 & 0.52 & 1.92 & 0.66 \\
  3 & 1.10 & 0.30 & 1.15 & 0.31 & 7.68 & 1.96 & 9.60 & 2.50 & 2.42 & 0.55 & 2.02 & 0.67 \\ 
  5 & 1.29 & 0.35 & 1.51 & 0.40 & 7.94 & 2.07 & 9.04 & 2.34 & 2.51 & 0.58 & 2.14 & 0.69 \\
  5 & 1.21 & 0.33 & 1.34 & 0.36 & 11.53 & 3.24 & 10.68 & 2.76 & 2.57 & 0.62 & 2.31 & 0.71 \\
  6 & 1.40 & 0.38 & 1.72 & 0.45 & 9.54 & 2.42 & 8.85 & 2.23& 2.83 & 0.70 & 2.61 & 0.78 \\
 7 & 1.54 & 0.42 & 1.97 & 0.51& 9.18 & 2.32 & 9.38 & 2.34 & 2.91 & 0.75 & 2.84 & 0.79 \\
  8 & 1.72 & 0.46 & 2.24 & 0.58 & 14.21 & 3.92 & 10.07 & 2.53 & 3.25 & 0.82 & 3.13 & 0.88 \\
  9 & 2.02 & 0.54 & 2.64 & 0.69 & 12.46 & 3.13 & 9.42 & 2.36 & 3.72 & 1.00 & 3.80 & 1.01 \\ 
  10 & 2.90 & 0.75 & 3.52 & 0.90 & 12.03 & 3.12 & 11.11 & 2.85 & 4.78 & 1.33 & 5.15 & 1.26 \\   \midrule
  Mean &  1.52 & 0.41 & 1.80 & 0.47 & 10.04 & 2.64 & 9.54 & 2.42 & 2.97 & 0.74 & 2.78  & 0.81\\
\bottomrule
\end{longtable}
\end{center} 

\vspace{-.5in}

Although the functional factor model introduced by \cite{Tavakoli2022} achieves the highest accuracy in terms of point forecasts among all the methods, the difference is not substantial when compared to our proposed FM-ANOVA.

Because of the two-stage FPCA, the method of \cite{GSY19} suffers information loss and can result in the worst accuracy among the three competitors. The maximum entropy model of \cite{PLC2019} was designed for modeling the age distribution of death counts. Hinging on the existence of finite statistical moments, it provides moderate point forecast accuracy.

\begin{figure}[!htb]
\centering
\includegraphics[width=0.47\textwidth]{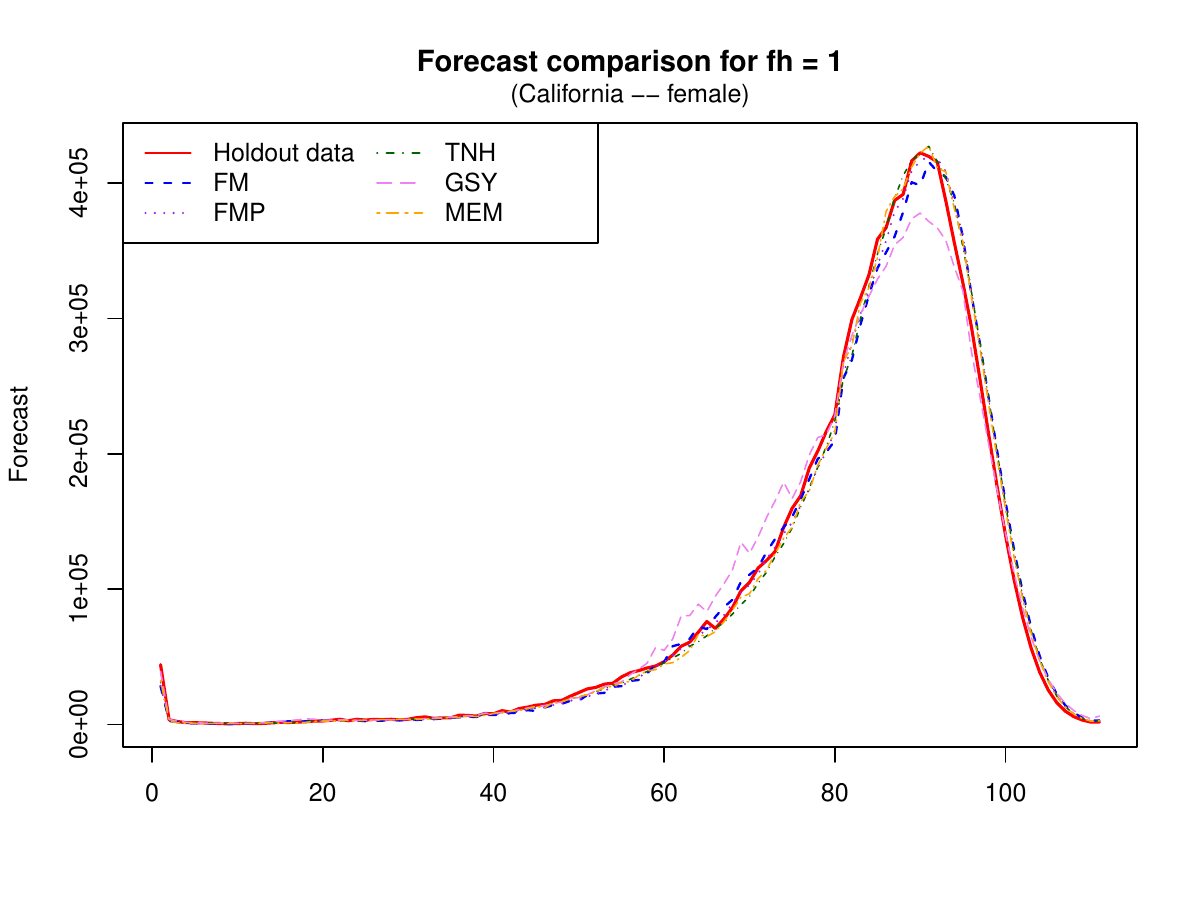}
\quad
\includegraphics[width=0.47\textwidth]{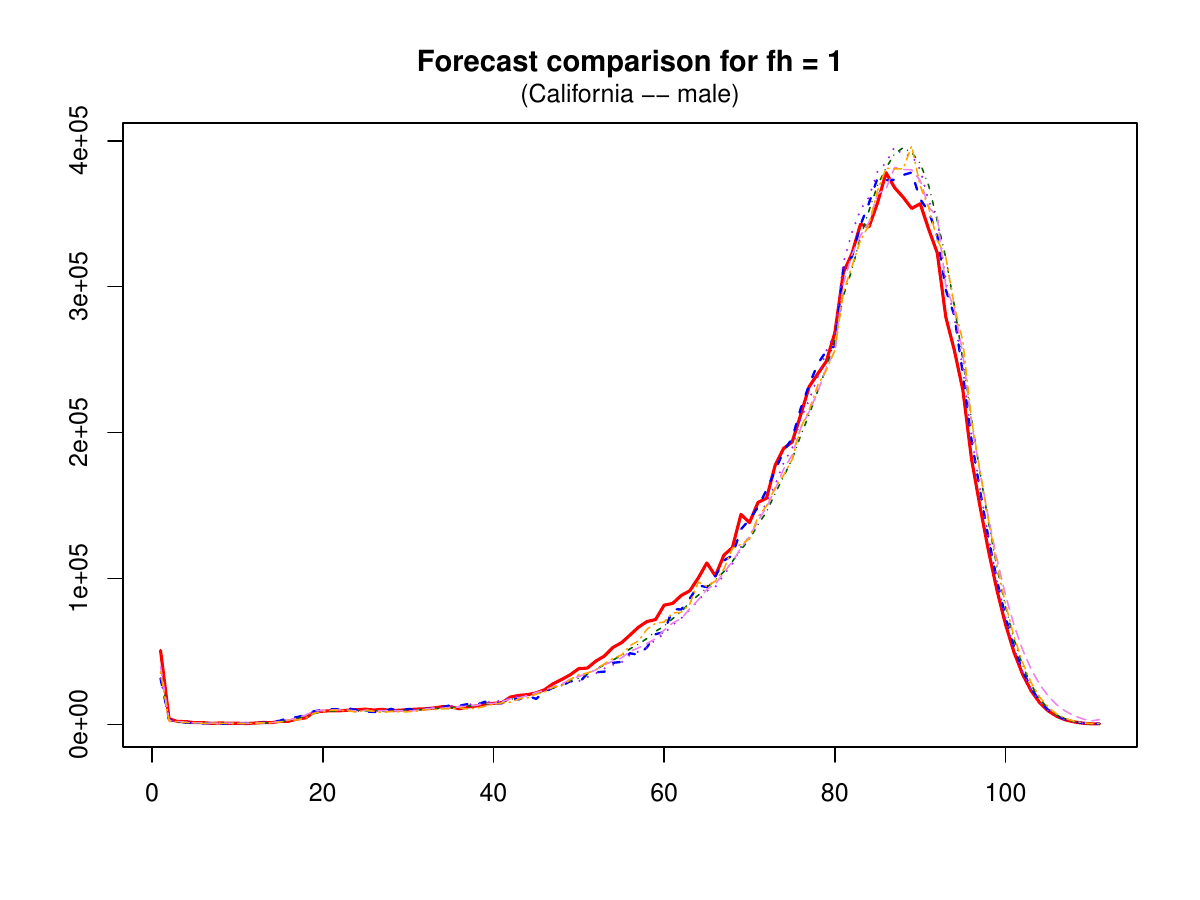}
\\
\includegraphics[width=0.47\textwidth]{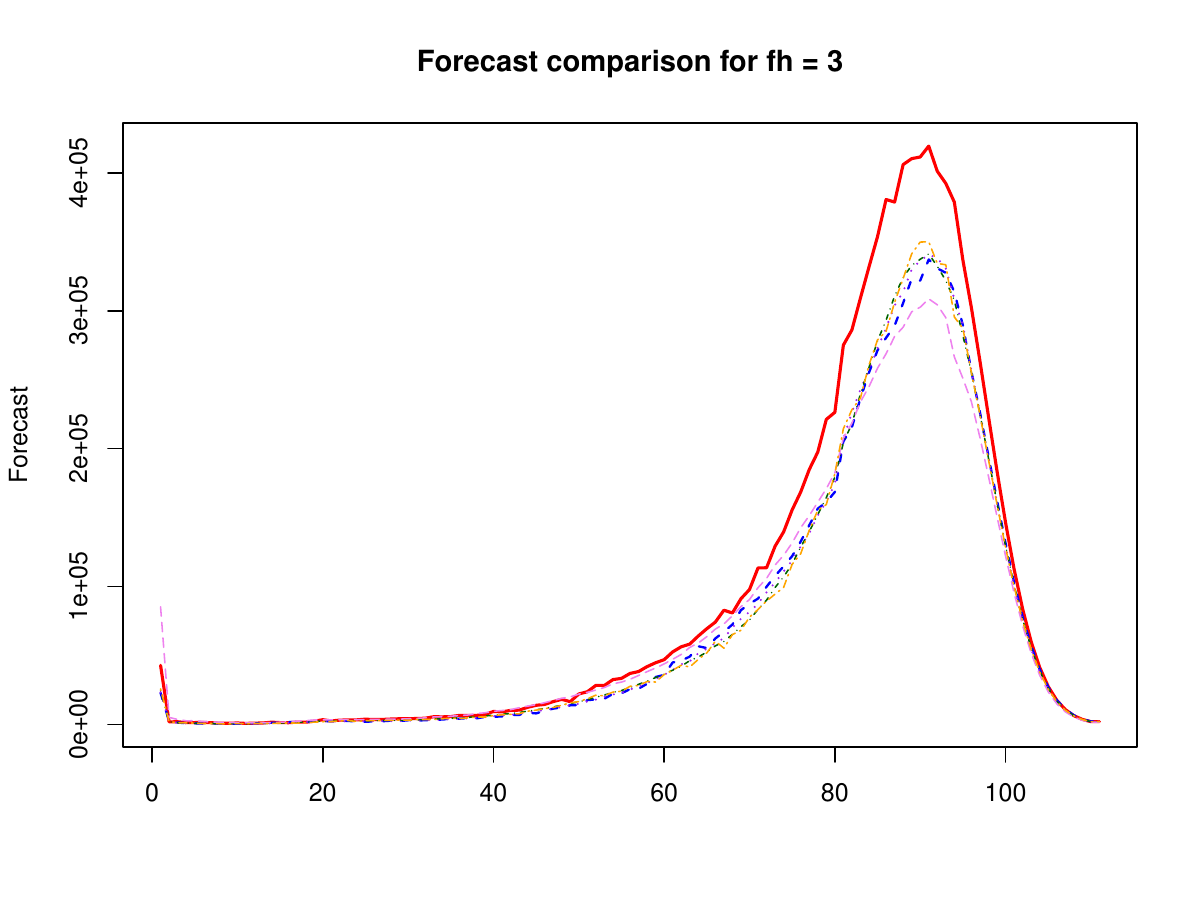}
\quad
\includegraphics[width=0.47\textwidth]{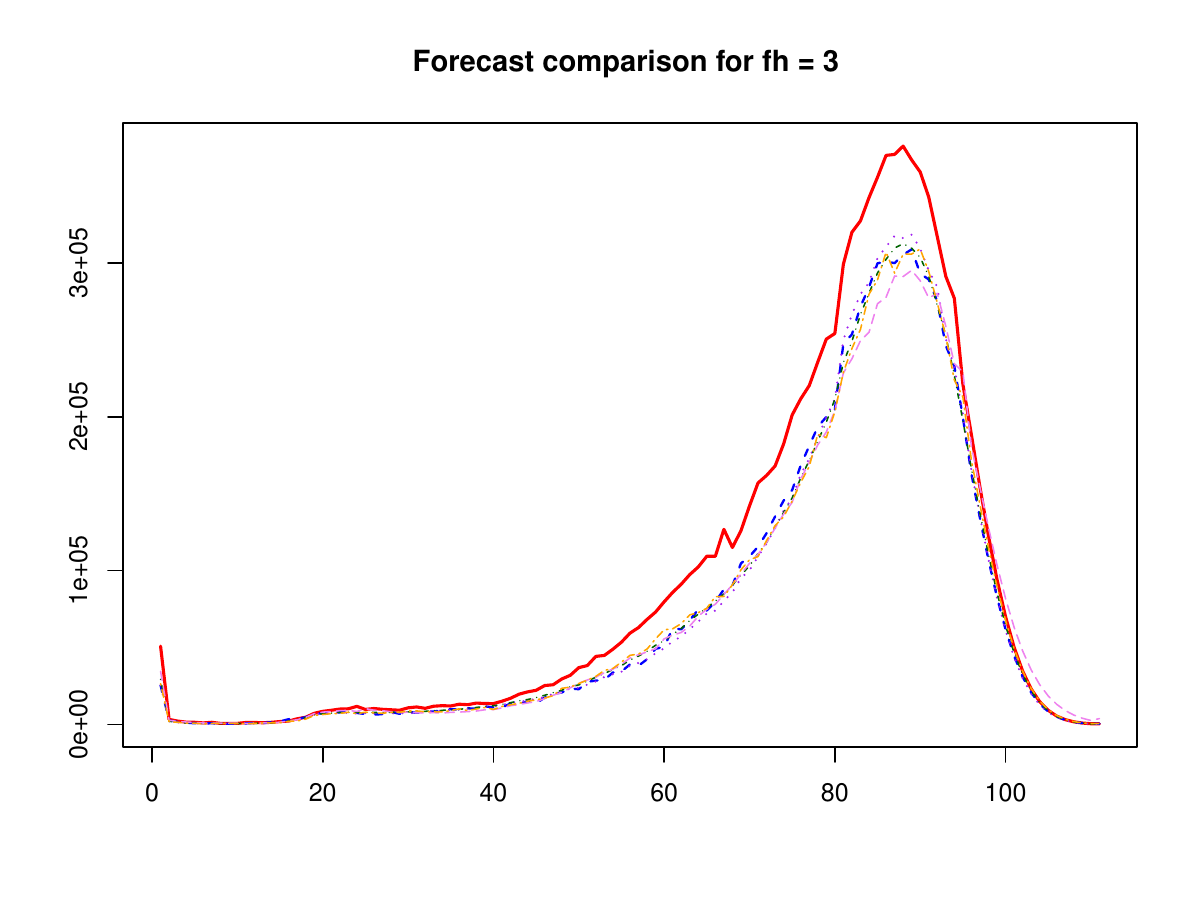}
\\
\includegraphics[width=0.47\textwidth]{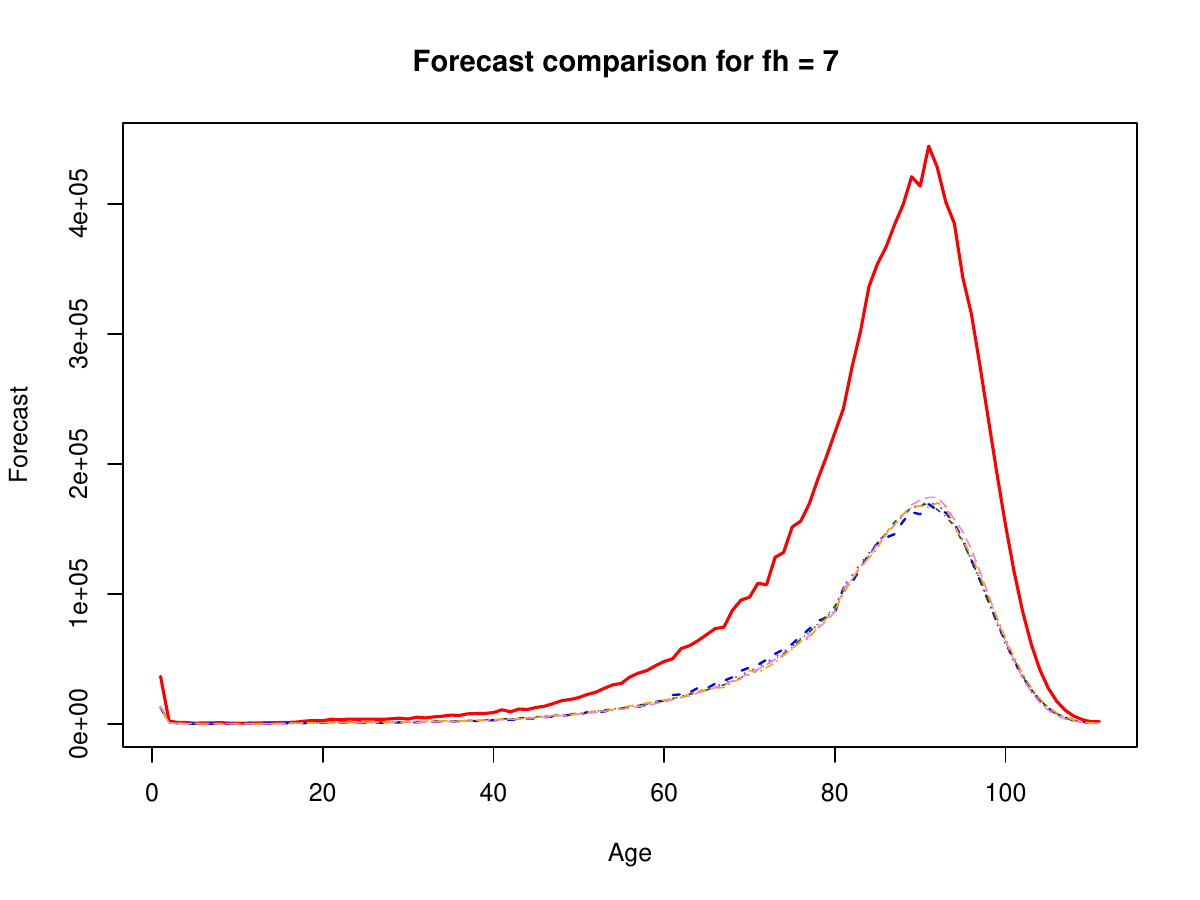}
\quad
\includegraphics[width=0.47\textwidth]{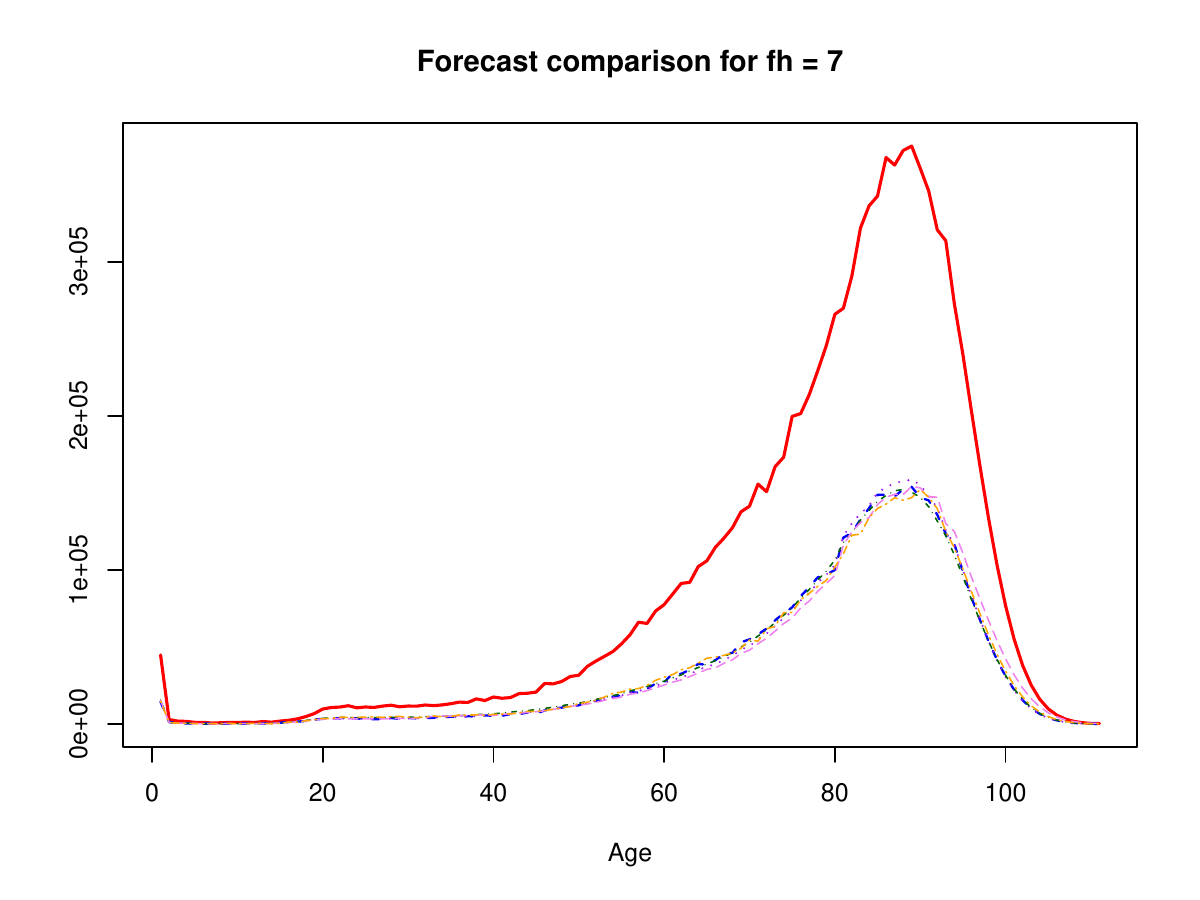}
\caption{Comparison of plots for the female and male data in  California with different horizons.}\label{fig:matrix-plots}
\end{figure}

To gain a clearer understanding of the forecast behavior of the proposed methods in comparison to the benchmark methods, Figure~\ref{fig:matrix-plots} provides an example analysis of the life-table death count data for the state of California, segmented by gender. The first column of the figure shows the forecast results for females, while the second column displays those for males. Each row represents a different forecast horizon, namely $h=1, 3, 7$. From these plots, it is evident that the proposed methods effectively capture the behavior of the holdout densities, particularly for the shorter forecast horizons ($h=1$ and $h=3$), where the predicted trends and density shapes closely align with the observed values. However, as the forecast horizon extends to $h=7$, we observe a noticeable decline in forecast accuracy for all methods, as anticipated, given the inherent challenges of comparably longer-term forecasting.

For further comparison, state- and gender-specific point forecast errors for all methods across multiple forecast horizons ($h=1, 2, \ldots, H$) can be explored interactively via the Shiny app: \url{https://cfjimenezv.shinyapps.io/Forecasting_density_valued_functional_panel_data/}.

\section{Conclusion}\label{sec:conclusion}

Understanding patterns in mortality across multiple populations is essential for local health policy decision-making. By modeling life-table death counts in multiple populations at the sub-national level, we expand the clr transformation to density-valued functional panel data. We apply the clr transformation to obtain unconstrained functional panel data. The time-varying components in the functional panel data are then obtained using two-way functional ANOVA decomposition across several states and genders.  In the unconstrained space, we apply a functional time series method to forecast the time-varying component. Using the inverse clr transformation, we obtain forecasts for the age-specific life-table death counts for multiple populations. The estimate of subpopulation mortality risks is needed to identify and understand the mortality patterns of vulnerable groups and track the effects of policy responses.

Within the clr transformation, we introduce the FMP-ANOVA and FM-ANOVA decompositions for density-valued functional panel data. Both approaches provide a decomposition to decompose the original HDFTS into deterministic and time-varying components within an unrestricted domain. We demonstrate that incorporating the clr transformation improves the forecast accuracy. 

The paper could be expanded in several ways, and we briefly mention three. 
\begin{inparaenum}
\item[1)] The HDFTS could also be extended beyond gender and state-specific populations in this research. Future extensions may include various subsets within each population (e.g., by socioeconomic class, ethnic group, or education level). 
\item[2)] In the case of outlying years, a robust clr transformation, such as that proposed by \cite{Filzmoser2009}, may be used. 
\item[3)] We outline as future work the development of uniform prediction bands and the assessment of empirical coverage probability difference for the life table death counts considered as densities.
\end{inparaenum}

\section*{Supplementary Materials}

\textbf{\Rlogo \ code for functional time series forecasting based on FMP-ANOVA and FM-ANOVA within the clr transformation.} Producing point forecasts from the two approaches described in the paper, including the clr transformation. The R codes are available at the following repository: 
\url{https://github.com/cfjimenezv07/CoDa_life_table_death_counts}\\
\textbf{\Rlogo \ code for shiny application.} Producing a shiny user interface for plotting every series and the results for point for the US life-table death count database.
The R codes are available at the following repository: 
\url{https://github.com/cfjimenezv07/CoDa_life_table_death_counts}

\section*{Acknowledgement}

We are grateful to the FDA workshop participants in Lille, France and the participants in the EcoSta2024 conference in Beijing, China. The first author thanks the financial support from  King Abdullah University of Science and Technology. The third author thanks fundings from Macquarie University Data Horizons consilience center, the Australian Research Council Discovery Project DP230102250 and Future Fellowship FT240100338.

\newpage
\bibliographystyle{chicago}
\bibliography{FPD}

\newpage
\appendix
\section{Number of functional principal components \texorpdfstring{$K=6$}{K=6}}\label{A1:K_6}
We conducted an additional analysis by fixing the number of retained components at $K=6$, as it has been shown that this number of components is enough for forecasting \citep{HBY13}. Table~\ref{tab:PFE_K} presents the results regarding the point forecast errors.

\begin{center}
\tabcolsep 0.58cm
\begin{longtable}{@{}lllllllll@{}}
\caption{\small With $K=6$, point forecast accuracy, as measured by the KLD and $\text{JSD}$, for the US between the FMP-ANOVA and FM-ANOVA. All values are multiplied by a factor of $100$.} \label{tab:PFE_K}\\
\toprule
	& \multicolumn{4}{c}{\textbf{FMP-ANOVA}} & \multicolumn{4}{c}{\textbf{FM-ANOVA}} \\
         & \multicolumn{2}{c}{\textbf{Female}} & \multicolumn{2}{c}{\textbf{Male}} & \multicolumn{2}{c}{\textbf{Female}} & \multicolumn{2}{c}{\textbf{Male}}  \\ \cmidrule{2-9}
        $h$ & \textbf{KLD}       & \textbf{JSD}    & \textbf{KLD}         & \textbf{JSD} & \textbf{KLD}       & \textbf{JSD}    & \textbf{KLD}         & \textbf{JSD}   \\ \midrule
\endfirsthead
\toprule
	& \multicolumn{4}{c}{\textbf{FMP-ANOVA}} & \multicolumn{4}{c}{\textbf{FM-ANOVA}} \\
         & \multicolumn{2}{c}{\textbf{Female}} & \multicolumn{2}{c}{\textbf{Male}} & \multicolumn{2}{c}{\textbf{Female}} & \multicolumn{2}{c}{\textbf{Male}}  \\ \cmidrule{2-9}
         $h$ & \textbf{KLD}       & \textbf{JSD$^g$}    & \textbf{KLD}         & \textbf{JSD$^g$} & \textbf{KLD}       & \textbf{JSD$^g$}    & \textbf{KLD}         & \textbf{JSD$^g$}   \\ \midrule
\endhead
\multicolumn{9}{r}{{Continued on next page}} \\ 
\endfoot
\endlastfoot
  1 & 2.08 & 0.62 & 2.10 & 0.59 & 1.18 & 0.34 & 1.27 & 0.34 \\ 
  2 & 2.02 & 0.60 & 2.11 & 0.58 & 1.16 & 0.33 & 1.34 & 0.36 \\ 
  3 & 2.04 & 0.59 & 2.29 & 0.62 & 1.25 & 0.35 & 1.50 & 0.40 \\ 
  4 & 2.13 & 0.61 & 2.50 & 0.67 & 1.39 & 0.39 & 1.67 & 0.44 \\ 
  5 & 2.30 & 0.66 & 2.71 & 0.72 & 1.55 & 0.43 & 1.86 & 0.49 \\ 
  6 & 2.53 & 0.72 & 3.03 & 0.80 & 1.78 & 0.49 & 2.13 & 0.55 \\ 
  7 & 2.67 & 0.75 & 3.35 & 0.89 & 1.97 & 0.53 & 2.39 & 0.62 \\ 
  8 & 2.91 & 0.80 & 3.74 & 0.99 & 2.26 & 0.61 & 2.75 & 0.71 \\ 
  9 & 3.40 & 0.93 & 4.30 & 1.13 & 2.89 & 0.79 & 3.26 & 0.84 \\ 
  10 & 4.53 & 1.20 & 5.72 & 1.49 & 4.07 & 1.05 & 4.58 & 1.17 \\ \midrule
Mean &   2.96 & 0.82 & 3.18 & 0.84 & 2.15 & 0.58 & 2.25 & 0.59   \\
\bottomrule          
\end{longtable}
\end{center}

\newpage
\section{Alternative methods without the clr transformation}\label{A2:wo_clr}

Table~\ref{tab:A2} displays the point forecast error results for two competitive methods: TNH by \cite{Tavakoli2022} and GSY by \cite{GSY19}. When these methods are directly applied, the generated functional curves may not belong to the functional space where the densities are located. Consequently, any negative values in the forecasts were set to zero, and the forecasts were subsequently normalized to $10^5$.

\begin{center}
\tabcolsep 0.227in
\begin{longtable}{@{}lllllllll@{}}
\caption{\small Point forecast accuracy, as measured by the KLD and $\text{JSD}$, for the US for alternative methods without the clr transformation. TNH refers to the method of \cite{Tavakoli2022}, GSY refers to the method of \cite{GSY19}. All values are multiplied by a factor of $100$.} \label{tab:A2}\\
\toprule
& \multicolumn{4}{c}{\textbf{TNH23}} & \multicolumn{4}{c}{\textbf{GSY19}} \\
& \multicolumn{2}{c}{\textbf{Female}} & \multicolumn{2}{c}{\textbf{Male}} & \multicolumn{2}{c}{\textbf{Female}} & \multicolumn{2}{c}{\textbf{Male}}  \\ \cmidrule{2-9}
$h$ & \textbf{KLD}       & \textbf{JSD}    & \textbf{KLD}         & \textbf{JSD} & \textbf{KLD}       & \textbf{JSD}    & \textbf{KLD}         & \textbf{JSD}   \\ \midrule
\endfirsthead
\toprule
& \multicolumn{4}{c}{\textbf{TNH23}} & \multicolumn{4}{c}{\textbf{GSY19}}  \\
& \multicolumn{2}{c}{\textbf{Female}} & \multicolumn{2}{c}{\textbf{Male}} & \multicolumn{2}{c}{\textbf{Female}} & \multicolumn{2}{c}{\textbf{Male}}  \\ \cmidrule{2-9}
$h$ & \textbf{KLD}       & \textbf{JSD$^g$}    & \textbf{KLD}         & \textbf{JSD$^g$} & \textbf{KLD}       & \textbf{JSD$^g$}    & \textbf{KLD}         & \textbf{JSD$^g$}   \\ \midrule
\endhead
\midrule
\multicolumn{9}{r}{{Continued on next page}} \\ 
\endfoot
\endlastfoot
 1 & 10.66 & 3.13 & 11.67 & 3.49 & 5.22 & 1.53 & 5.53 & 1.55 \\ 
  2 & 10.68 & 3.14 & 11.74 & 3.52 & 5.99 & 1.77 & 5.47 & 1.53 \\ 
  3 & 10.98 & 3.23 & 11.84 & 3.54 & 6.67 & 1.99 & 5.57 & 1.54 \\ 
  4 & 11.33 & 3.33 & 11.87 & 3.55 & 6.12 & 1.82 & 6.59 & 1.81 \\ 
  5 & 11.61 & 3.41 & 11.91 & 3.57 & 6.50 & 1.92 & 7.69 & 2.18 \\ 
  6 & 12.09 & 3.56 & 11.94 & 3.58 & 8.03 & 2.41 & 7.95 & 2.23 \\ 
  7 & 12.70 & 3.74 & 12.08 & 3.62 & 7.54 & 2.25 & 8.84 & 2.46 \\ 
  8 & 13.39 & 3.95 & 12.32 & 3.70 & 8.27 & 2.46 & 11.23 & 3.23 \\ 
  9 & 14.05 & 4.16 & 13.01 & 3.93 & 8.97 & 2.66 & 9.49 & 2.75 \\ 
  10 & 15.40 & 4.60 & 14.58 & 4.45 & 10.89 & 3.26 & 19.44 & 5.91 \\  
     \midrule
  Mean &   12.29  & 3.63  &  12.30 & 3.70  & 7.42 & 2.21 &  8.78 & 2.52 \\
\bottomrule
\end{longtable}
\end{center}

\end{document}